\documentstyle[12pt]{article}
\title{\large\bf High $E_T$ jets at $p \bar p$ collisions \\
and triple gauge boson vertex}
\author{ 
B.A.Arbuzov  \\
\small\it{Institute for High Energy Physics, Protvino,}\\
\small\it {Moscow region, 142284, Russia}} 
\date{ }
\begin{document}
\maketitle

{\bf Abstract}
\bigskip

{\small Triple gauge boson vertex, which is inherent to 
the model of a dynamical breaking of the electroweak symmetry, 
leads to an effective four-fermion interaction of quarks. 
Calculation of the effective coupling constant in the framework of 
the model results in agreement with recent information on 
enhancement of inclusive high $E_T$ jet cross section at $\sqrt{s} = 1.8\,TeV$. 
Thus this effect may be connected with triple gauge boson vertex with  
coupling constant $\lambda \approx 0.7$, which will clearly manifest itself 
at the soon forthcoming LEP II. }
\bigskip

PASC: 12.15; 14.80.Er

\bigskip
\bigskip

In the recent Fermilab experiment [1] the effect of an enhanced cross section 
of inclusive jets production at $\sqrt{s} = 1.8\,TeV$ with $E_T > 200\,GeV$ 
is obtained. The authors state, that this effect could be connected with 
consequences of quark compositeness, leading e.g. to effective four-fermion 
interaction [2]
\begin{equation}
L_{eff}\,=\,\pm\,\frac{4\,\pi}{2 \,\Lambda_C^2}\,\bar q_L \gamma_\mu q_L\,
\bar q_L \gamma_\mu q_L\,;
\end{equation}
where $q$ corresponds to light quark pair $u,\;d$ and currents in 
interaction (1) are isoscalar. The best agreement of data [1] with (1) 
is achieved for
\begin{equation}
\Lambda_C\;=\;1.6\;TeV\;.
\end{equation}

In this letter we would point out, that effective four-fermion interaction 
appears not only in the case of quarks substructure but also in the 
case of an existence of triple gauge vertex 
($\sim\,\epsilon^{a b c}W_{\mu \nu}^a\,W_{\nu \rho}^b\,W_{\rho}^c$), 
which appears in a selfconsistent way in the model of a dynamical 
breaking of the electroweak symmetry [3 -- 6]
\begin{eqnarray}
& &\Gamma_{\mu \nu \rho}^{a b c} (p,\,q,\,k)\,=\,\frac{g\,\lambda}{M_W^2}\,
\epsilon^{a b c}\,
F(p,\,q,\,k)\,\Bigl(g_{\mu \nu}(p_\rho (qk) - q_\rho (pk))\,+\nonumber\\
& &+\,g_{\nu \rho}(q_\mu (pk) - k_\mu (pq))\,+
\,g_{\rho \mu} (k_\nu (pq) - p_\nu (qk))\,+\,k_\mu p_\nu q_\rho\,
-\,q_\mu k_\nu p_\rho\,\Bigr)\,;
\end{eqnarray}
where formfactor 
$F$ restricts the small momenta region of action of the vertex by 
an effective cut-off $\Lambda_{EW}$, which is of few $TeV$ order 
of magnitude. According to estimates [5], [6], which are consistent 
with $t$ - quark production cross section, 
$Z\rightarrow \bar b b$ width and experimental limitations [7], [8]
\begin{equation}
\Lambda_{EW}\,\approx \,4.5\,TeV\,;\qquad \lambda\,\approx\,0.7\,.
\end{equation}
Note, that additional gauge boson vertices are repeatedly introduced 
basing on various arguments (see e.g. [9]). 

Let us calculate effective four-fermion terms in one loop approximation 
using vertex (3) and usual EW interaction
\begin{equation}
\frac{g}{2}\,\biggl(\bar q_L\,\gamma_\mu\,\tau^a\,q_L\,+\,
\sum_j\,\bar Q_L^j\,\gamma_\mu\,\tau^a\,Q_L^j\biggr)\;W^a_\mu\,;
\end{equation}
where $j\,=\,1,\,2$ marks generations of heavier quarks. We use here 
the electroweak interaction without symmetry breaking, because we deal 
with $TeV$ energy region, corresponding to restoration of the symmetry. 

We have two types of diagrams: $s$ - type and $t$ - type. They both 
contain triangle diagram with one vertex (3) and two vertices (5). 
Calculation of this diagram gives the following result (see also [10])
\begin{equation}
\frac{\lambda\,g^3}{64\,\pi^2\,M_W^2}\,ln\biggl(\frac{\Lambda_{EW}^2}{M_W^2}
\biggr)\;\bar q_L\,\tau^a\,(\gamma_\mu k^2 - k_\mu \hat k)\,q_L\,W_\mu^a\,;
\end{equation}
where $k$ is the momentum of $W$. The second term in brackets is proportional 
to $M_q$, so it is negligible. The first one is proportional to $k^2$, so 
being multiplied by $W$ propagator $(k^2 - M_W^2)^{-1}$ it leads to 
contact four-fermion interaction for sufficiently large $k^2 \gg M_W^2$ . 
We have two $s$ - type diagrams (with triangle to the left and to the right) 
and two $t$ - type ones. Now we obtain the effective four-fermion 
interaction, of which we write down terms, containing at least one 
$\bar q\,q$ factor and thus giving contribution to $\bar p p$ processes
\begin{eqnarray}
& &L_{eff}\,=\,\frac{G}{2}\biggl\{\bar q_L\,\gamma_\mu\,\tau^a\,q_L\,
\biggl(\bar q'_L\,\gamma_\mu\,\tau^a\,q'_L\;+\;2\,\sum_j\,\bar Q_L^j\,
\gamma_\mu\,\tau^a\,Q_L^j\,\biggr)\;+\nonumber\\
& &+\;\bar q_L\,\gamma_\nu\,\tau^b\,q'_L\;\bar q'_L\,\gamma_\nu\,
\tau^b\,q_L\,\biggr\}\;;\\
& &G\,=\,\frac{\lambda\,\alpha_W^2}{M_W^2}\,ln\biggl(\frac
{\Lambda_{EW}^2}{M_W^2}\biggr)\;;\qquad \alpha_W\,=\,\frac{g^2}{4\,\pi}
\,=\,\frac{\alpha}{\sin^2\theta_W}\;.
\nonumber
\end{eqnarray}
Here we introduce primes to distinguish initial (not primed) and final 
(primed) light quarks in our diagrams. Really the last term in Eq.(7) 
corresponds to $t$ - type diagrams and it is convenient to perform 
Fiertz transformation 
\begin{equation}
\bar q_L \gamma_\nu \tau^b q'_L\,\bar q'_L \gamma_\nu \tau^b q\,=\,
\frac{3}{2}\,\bar q_L \gamma_\nu q\,\bar q'_L \gamma_\nu q'_L\,-
\,\frac{1}{2}\,\bar q_L \gamma_\nu \tau^a q\,\bar q'_L 
\gamma_\nu \tau^a q'_L\;.
\end{equation}
From Eqs. (7) and (8) we have final form of the effective interaction
\begin{eqnarray}
& &L_{eff}\,=\,\frac{G}{2}\,\biggl(\,\frac{3}{2}\,\bar q_L\,\gamma_\mu\,
q_L\,\bar q_L\,\gamma_\mu\,q_L\,+\,\frac{1}{2}\,
\bar q\,\gamma_\mu\,\tau^a\,q_L\,\bar q_L\,
\gamma_\mu\,\tau^a\,q_L\,+\nonumber\\
& &+\,2\,\bar q\,\gamma_\mu\,\tau^a\,q_L\,
\sum_j\,\bar Q_L^j\,\gamma_\mu\,\tau^a\,Q_L^j\biggr)\,.
\end{eqnarray}
The first term in lagrangian (9) coincides with Eq.(1) provided 
\begin{equation}
\Lambda_C\,=\,\sqrt{\frac{8\,\pi}{3\,G}}\,=\,\frac{2\,M_W}{\alpha_W}\,
\sqrt{\frac{\pi}{3\,\lambda\;ln\bigl(\Lambda_{EW}/M_W\bigr)}}\;.
\end{equation}

Substituting values (4) into Eq.(10) we have $\Lambda_C\,=\,2.87\,TeV$ , 
that is not so far from experimental value (2). 

However other terms in Eq.(9) also give contribution to the cross 
section. For example, we may start reasoning in the following way. 
Interference terms with leading QCD diagrams, which 
are describing the main effect of experiment [1] in the region 
$200\,GeV\,< E_T\,<\,300\,GeV$ , 
are caused by the first two terms of Eq.(9) and the contribution of the 
second term is equal to that of the first one. This result is due to 
simple relation
\begin{displaymath}
Trace\Bigl(\tau^a\,\tau^a\Bigr)\,=\,3\,Trace\Bigl(\,I\,\Bigr)\,.
\end{displaymath}
Therefore the result for the effective $\Lambda_C$ has to be divided by 
$\sqrt{2}$ (amplitudes of production of heavy quarks do not interfere 
with QCD terms)
\begin{equation}
\Lambda_C\,=\,\frac{2.87\;TeV}{\sqrt{2}}\,=\,2.03\,TeV\,.
\end{equation}
Result (11) is really close to value (2) and agrees with 
data [1] within their uncertainties. Let us mention, that the boundary 
value of restrictions [8] $\lambda = 0.85$ leads instead of (11) 
to $\Lambda_C = 1.84\,TeV$ , that almost coincides with value (2).

To conclude we would state that exciting results [1] 
could be interpreted not only in terms of quark compositeness, but also 
in terms of anomalous gauge vertices, which also lead to an effective 
four-fermion interaction. Parameter $\lambda$ in this case has to be 
sufficiently large, just on the boundary of experimental limitations 
[7], [8]. Such "strong" triple gauge vertex without doubt can be detected 
at the soon forthcoming LEP II facilities in the reaction of $W$ pair 
production, so the way to verify the variant being discussed here is 
quite straightforward.

The present work is partially supported by the Russian Foundation of 
Fundamental Researches under project 95-02-03704.

\end{document}